# Open-MPI *over* MOSIX

**Parallel Computing in a Clustered World**

Adam Lev-Libfeld*, Supervised by: Amnon Barak† and Alex Margolin‡
School of Computer Science and Engineering
The Hebrew University of Jerusalem, Israel

September 14, 2009

**Abstract**

Recent increased interest in Cloud computing emphasizes the need to find an adequate solution to the load-balancing problem in parallel computing - efficiently running several jobs concurrently on a cluster of shared computers (nodes). One approach to solve this problem is by preemptive process migration- the transfer of running processes between nodes. A possible drawback of this approach is the increased overhead between heavily communicating processes. This project presents a solution to this last problem by incorporating the process migration capability of MOSIX into Open-MPI and by reducing the resulting communication overhead. Specifically, we developed a module for direct communication (DiCOM) between migrated Open-MPI processes, to overcome the increased communication latency of TCP/IP between such processes. The outcome is reduced run-time by improved resource allocation.

This document is *Open Source Secure*:
Written from scratch in Emacs on a Unix machine using LaTeX *version* $2_\varepsilon$

*adamlev@cs.huji.ac.il, ID 200226983, Corresponding author.
†amnon@cs.huji.ac.il
‡alexam02@cs.huji.ac.il

*To Dad*

# Acknowledgments

Special thanks to *Amnon Shiloh* for the extensive help in all MOSIX related issues.

Thanks to *Tal Maoz, Tal Ben-Nun and the System group* for the moral and technical support.

All graphs, graphics, diagrams and charts, as well as any other artworks in this document is the result of the original work of the author unless mentioned otherwise. That said, I whould like to thank *Tuvia Lev* for his assistance in the graphical editing stage of this paper.

## Thank you all!

# Contents

# List of Figures

# Summary

Recent increased interest in Cloud computing emphasizes the need to find an adequate solution to the load-balancing problem in parallel computing - efficiently running several jobs concurrently on a set (cluster) of shared computers (nodes).One approach to solve this problem is by using preemptive process migration- the transfer of running processes between nodes. A possible drawback of this approach is the increased overhead between heavily communicating processes, which are a typical attribute of Open-MPI (OMPI) processes.

This project presents a solution to this last problem by incorporating the preemptive process migration and automatic load balancing capabilities of the MOSIX into OMPI and by reducing the resulting communication overhead.

Specifically, we developed a module for direct communication (DiCOM) between migrated OMPI processes, to overcome the increased communication latency of TCP/IP between such processes. The outcome is reduced run-time by improved resource allocation.

Since our main goal was to achieve local (non-migrated) TCP/IP net performance (or very close to it) in both local and migrated processes and after inspecting the source code of OMPI, we decided that due to it's modular structure it will be best to implement the MOSIX Direct Communication (DiCOM) module in the same form of the TCP/IP protocol, including *sockets,* and the function: *connect, accept, send, recv* etc., based on a variant of the existing, highly optimized OMPI TCP module.

The main advantages of this approach are to enable an easy integration of our module into any future TCP/IP-based applications or toolkits, and also allow comparisons between the net performance of DiCOM and TCP/IP, from these comparisons we concluded (as can be seen in Figure 1):

1. The performance of TCP/IP between **non-migrated** process is slightly better then DiCOM for all message sizes. The average slowdown is about **17%** for all message sizes. This is due to the fixed overhead of the DiCOM protocol.
2. DiCOM is slightly slower than **migrated** TCP/IP for small message sizes (below 32k). For larger message sizes, DiCOM is increasingly better than TCP/IP between migrated processes, with an average improvement of **52%** for all of the measured message sizes.

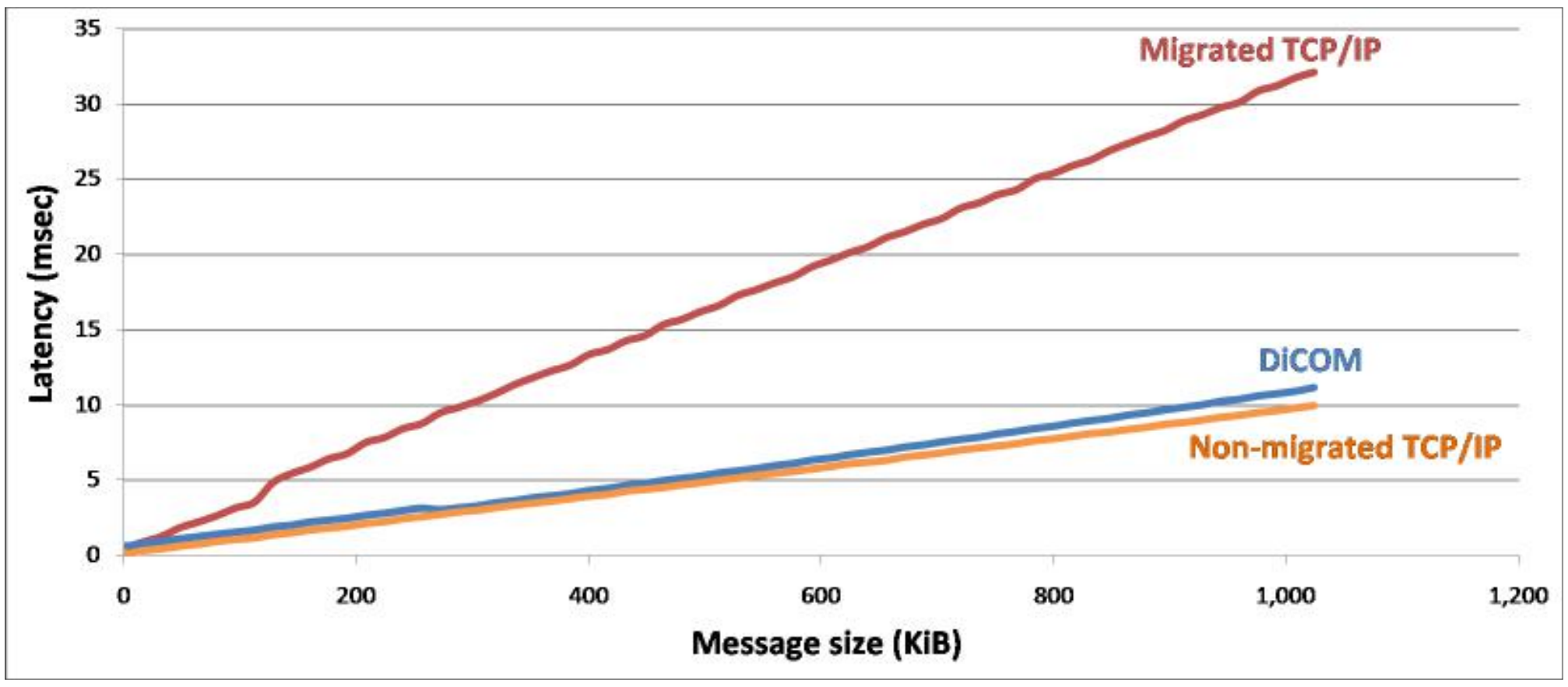

Figure 1: DiCOM compared to TCP/IP
TCP send/recv latency with and without migration (red and orange respectively) and DiCOM send/recv latency (blue) [mSec] as a function of message size [KiB].

# Chapter 1

# Introduction

## 1.1 Clusters, Grids & Clouds

With the increasing need for ever more powerful yet cost efficient computers, the search for better performance had led researchers of High-Performance Computing (HPC) to try and combine the computing power of several (cluster) computers (nodes) to increase total throughput, and think of new ways to utilize more of the existing computing resources in their organization, thus increasing the resource usage efficiency. Concepts like **NOW**[1], **LOBOS Cluster** [2] and **Grid** [3] are being formed, and as the technology advanced, its economic sense appealed to investors and entrepreneurs around the world.

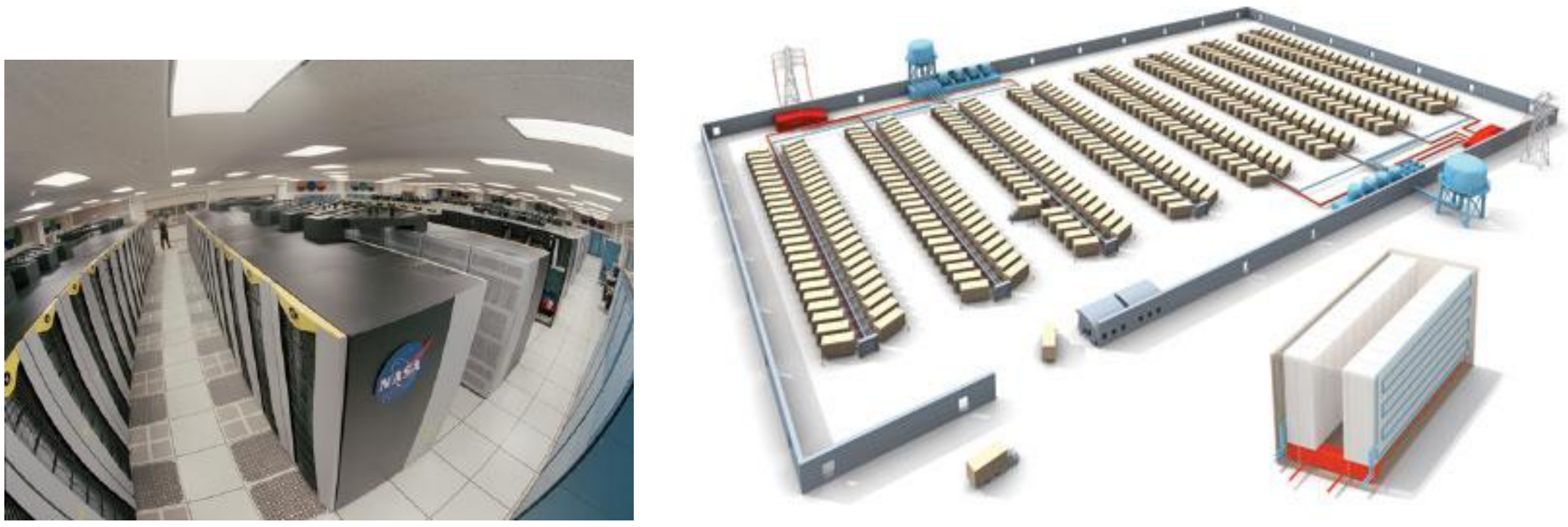

Figure 1.1: Computer clusters, now and in the near future
**Right:** A computer cluster owned by NASA. Source: Wikipedia. **Left:** The Million-Server Data Center (computer and storage Cluster), a near future. Source: IEEE Spectrum [6].

Today some of the biggest names and best known corporations of the computer world are involved in the Cluster Computing (mostly **Cloud Computing**[4]) among them are: Google, Yahoo, Sun, Microsoft, Amazon and more. This gives a major financial motivation to research the manners to make the most of the huge (tens of thousands of multi-core servers) Clusters owned by these firms and others.

## 1.2 Distributed Computing

Distributed Computing can be shortly described as the usage of a Cluster (or Grid) containing (sometimes various) processing element and hardware components to concurrently perform (run) multiple processes, programs or jobs under a controlled regime[5].

Distributed Computing systems often have some optimization mechanism that enables better use (more efficient in terms of time and cost) of the given resources. This optimization mechanism is usually some kind of a load balancing algorithm, in most cases this is a "terminate and restart"[6] mechanism, but in this project we use the **MOSIX** Distributed Computing system which uses the **preemptive process migration** load balancing method, the transparent[7] transfer of running processes between nodes. MOSIX also contains an internal information dissemination (gossip) mechanism and various algorithms aimed to optimize the usage of the systems clusters.

[1] Network Of Workstations, sometimes called COW – Cluster Of Workstations

[2] Lots Of Boxes On Shelves, refers to the usage of a cluster of "regular" low-to-middle-end desktop computers rather then specialized high-end HPC hardware, which is referred to as just a "Cluster"

[3] Several Clusters working together, same as Multi-Cluster. Grid is sometimes mistaken with other types of Clusters due to its constituent of various discrete elements.

[4] Cloud computing - a style of computing in which dynamically scalable and often virtualized resources are provided as a service over the Internet. Also defined as a computing paradigm where the boundaries of computing will be determined by economic rationale rather than technical limits by Prof. Ramnath K. Chellappa in ref [3]

[5] Set of rules, conditions and limitations that state the ability of an element of the system (either hardware or software) to perform and action or receive a service (such as when to migrate).

[6] Terminate the process and restart it later elsewhere. Used by most of the commercial Distributed Computing systems due to the fact that the usual work load consists of many discrete extremely small jobs (think Google search jobs - tiny, but in the millions). This method is not functional for Clusters on which job arrival is slow, bursty and the arriving jobs are of relatively large size, such as universities and other research facilities.

[7] Meaning the process is unaware of the transfer being more then a context switch.

## 1.3 Parallel Computing

Parallel Computing is the usage of multiple, concurrent processes to achieve a collective, united goal (operation) which is the result of the combined labor of the individual processes. In this computing paradigm the processes are either intercommunicating or synchronized, so, understandably, it is not possible to use the "terminate and restart" system to balance the load, since the early termination of any of the processes may puts the entire operation at risk of failure in the case of intercommunicating processes, and in any case will significantly delay the conclusion of the operation. In this project we worked with Open-MPI (OMPI) a vastly used open source MPI-2[8] implementation (see Appendix for more details about MPI and OMPI).

## 1.4 Issues concerning parallel (OMPI) applications performance

### 1.4.1 Without preemptive process migration load balancing

As mentioned before, the lack of load balancing mechanisms may lead to inefficient allocation of resources (I.e. CPU time) and the undesirable situation when the slowest instance (the result of either a busier or weaker node) is causing the entire job to slow down (see Figure 1.2). The MOSIX system prevents such situations by spreading the computational resources evenly among the MPI instances, thus allowing a fair distribution of resources and better performance for the entire job.

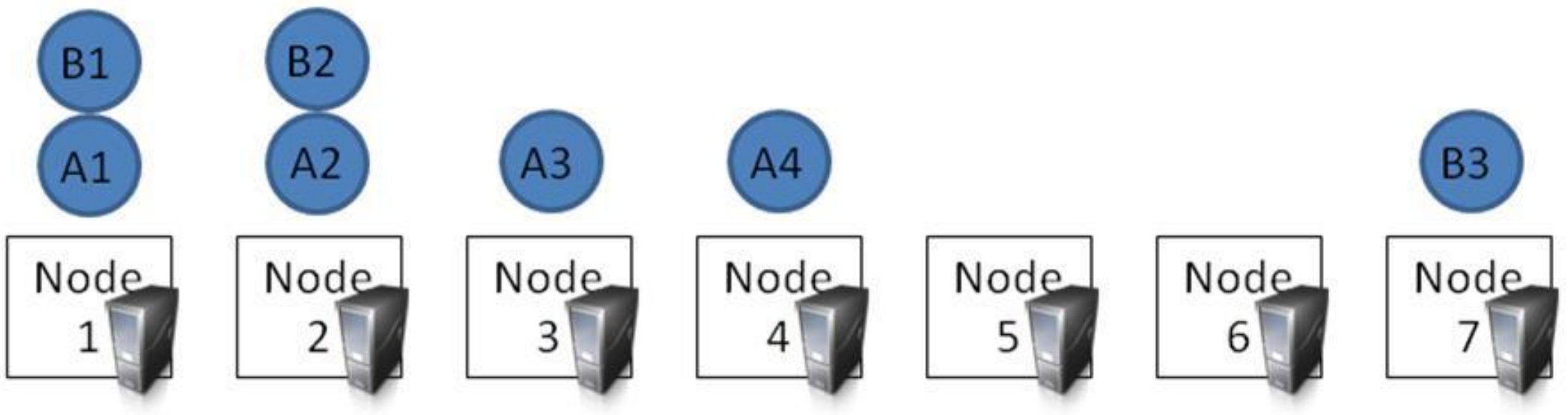

Figure 1.2: Load imbalance example
Due to imbalanced load both job A and job B are slowed down, migration of processes from nodes 1 and 2 to nodes 5 and 6 will allow better performance to all the jobs.

### 1.4.2 When introducing MOSIX preemptive process migration

Normally, MOSIX processes do all their I/O (and most system-calls) via their home-node. This can be slow since operations are limited by the network speed and latency and obviously inefficient due to the fact that the "detour"(see Figure 1.4) in the communication path. Moreover, rerouting the communications of all the migrated processes through the home-node causes a non-linear increased load in that node (see Figure 1.3), slowing all the running processes.

[8] Message Passing Interface, version 2

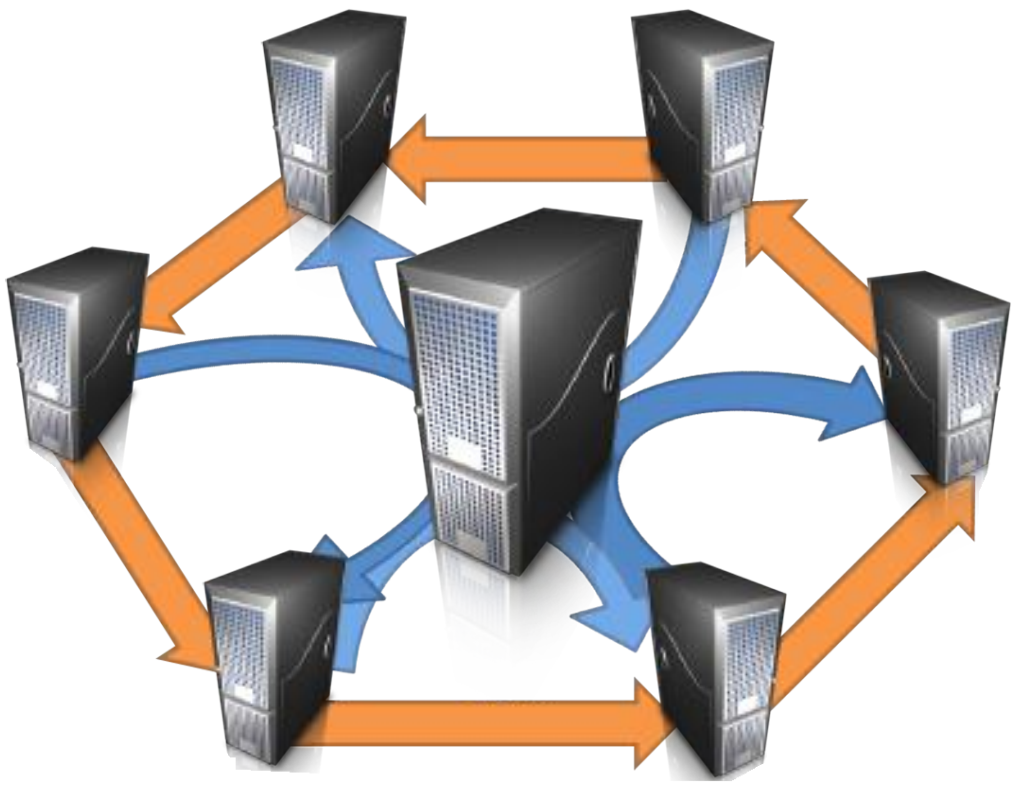

Figure 1.3: DiCOM and TCP/IP in a ring topology
Migrated TCP/IP(blue) and DiCOM (orange) connections in a ring topology, with the central node as the home node and all the processes are individually migrated to the outer ring of nodes.

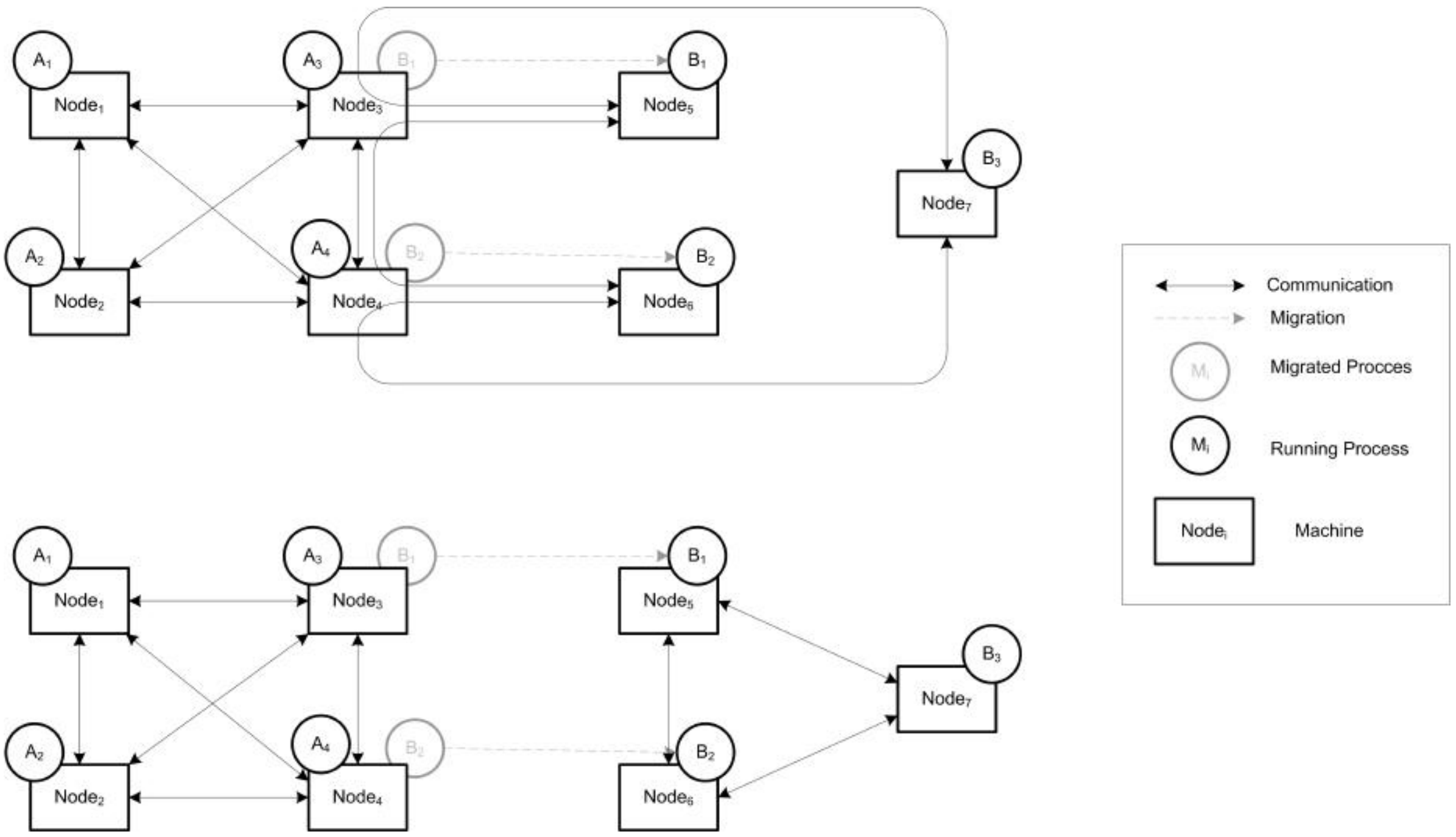

Figure 1.4: Network view before and after introducing DiCOM
Network view before (above) and after (below) introducing DiCOM to a 7 node cluster with a mesh (all to all) topology.

## 1.5 Project objective & requirements to determine success

Implement a TCP/IP compatible protocol, which uses the built in, gossip information dissemination algorithm of MOSIX, to allow migrated processes to communicate directly, without using the "home-node" of each process as a mediator, improving the efficiency of the system. The proposed protocol will be able to distribute network load evenly throughout the working cluster and minimize communication path resulting in proved ability to out-perform the state of the art (normal TCP/IP connection) without changes to the user code.

# Chapter 2

# Methods and Materials

## 2.1 Methods, Concepts and Algorithms

### 2.1.1 Path Optimization and Network Load Distribution

As the basic idea behind the entire project, path optimization will minimize the number of nodes engaged in any data transaction, allowing the system to achieve a more balanced network load throughout the working Cluster.

Currently, both MOSIX and OMPI does not provide any path optimization mechanisms: For example (Figure 2.1), if process X, whose home-node is H1, is running on node R1, wishes to send a message over a socket to process Y, whose home-node is H2 and is running on node R2, then without DiCOM the message has to pass through the following path:

R1 -> H1 -> H2 ->R2 (marked blue in the Figure 2.1).

Using DiCOM (orange in Figure 2.1), the message will pass directly from R1 to R2. Moreover, if X and Y run on the same node, communication done via shared memory and the network is not used at all.

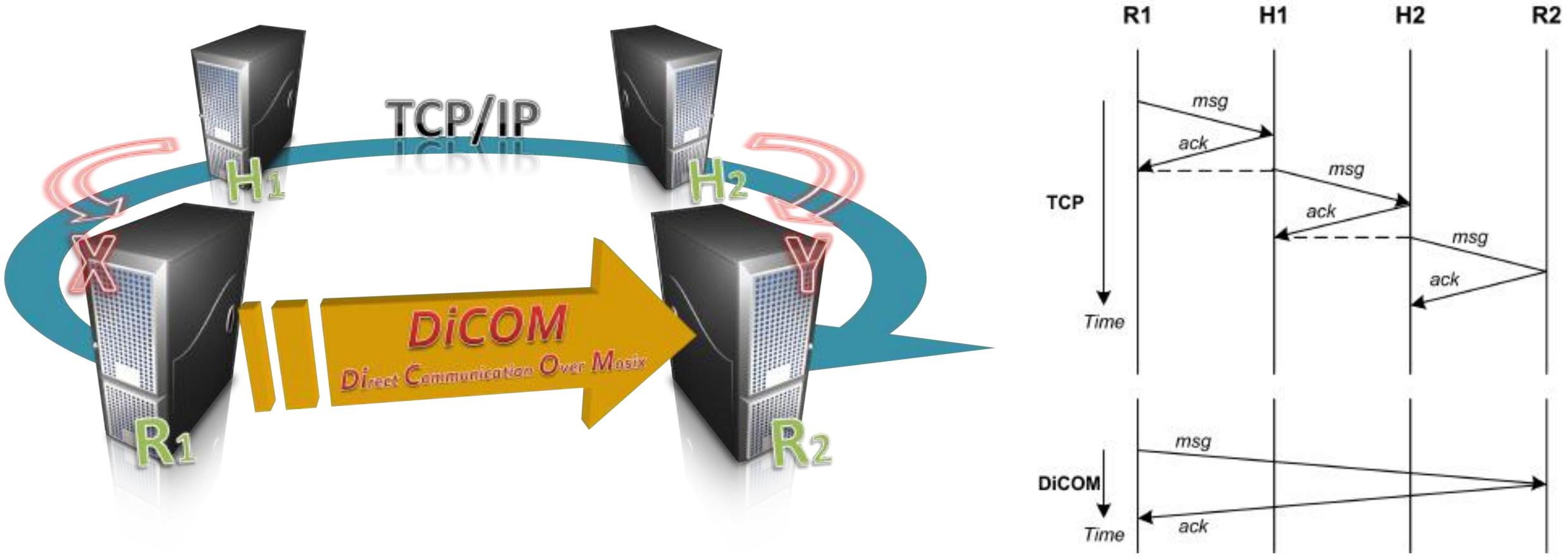

Figure 2.1: TCP/IP and DiCOM communication
TCP/IP (blue) and DiCOM (orange) communication path in a migrated configuration and the appropriate protocol diagram.

The use of path optimizing mechanisms will shorten the communication path, improving latency and net fault rate.

### 2.1.2 Design concepts

Since this project is part of both the MOSIX and the OMPI projects, it is necessary that the solution will be as simple and as elegant as possible. After inspecting the source code of OMPI, we decided that due to it's modular structure[1] it will be best to implement the MOSIX Direct Communication (DiCOM) module in the same form of the TCP/IP protocol, including *sockets , bind, connect, accept, select, close, send, recv* etc., based on a variant of the existing, highly optimized OMPI TCP module (Figure 2.2).

Effectively, most of the DiCOM code is in the MOSIX source code and the DiCOM module in OMPI is a set of functions compatible with the TCP/IP interface but using the DiCOM functions within MOSIX. The main advantages of this approach are to allow comparisons between the net performance of DiCOM and TCP/IP, and also enable an easy integration of our module into any future TCP/IP-based applications or toolkits.

[1] For further reading about the modular structure of OMPI refer to the appropriate Appendix section.

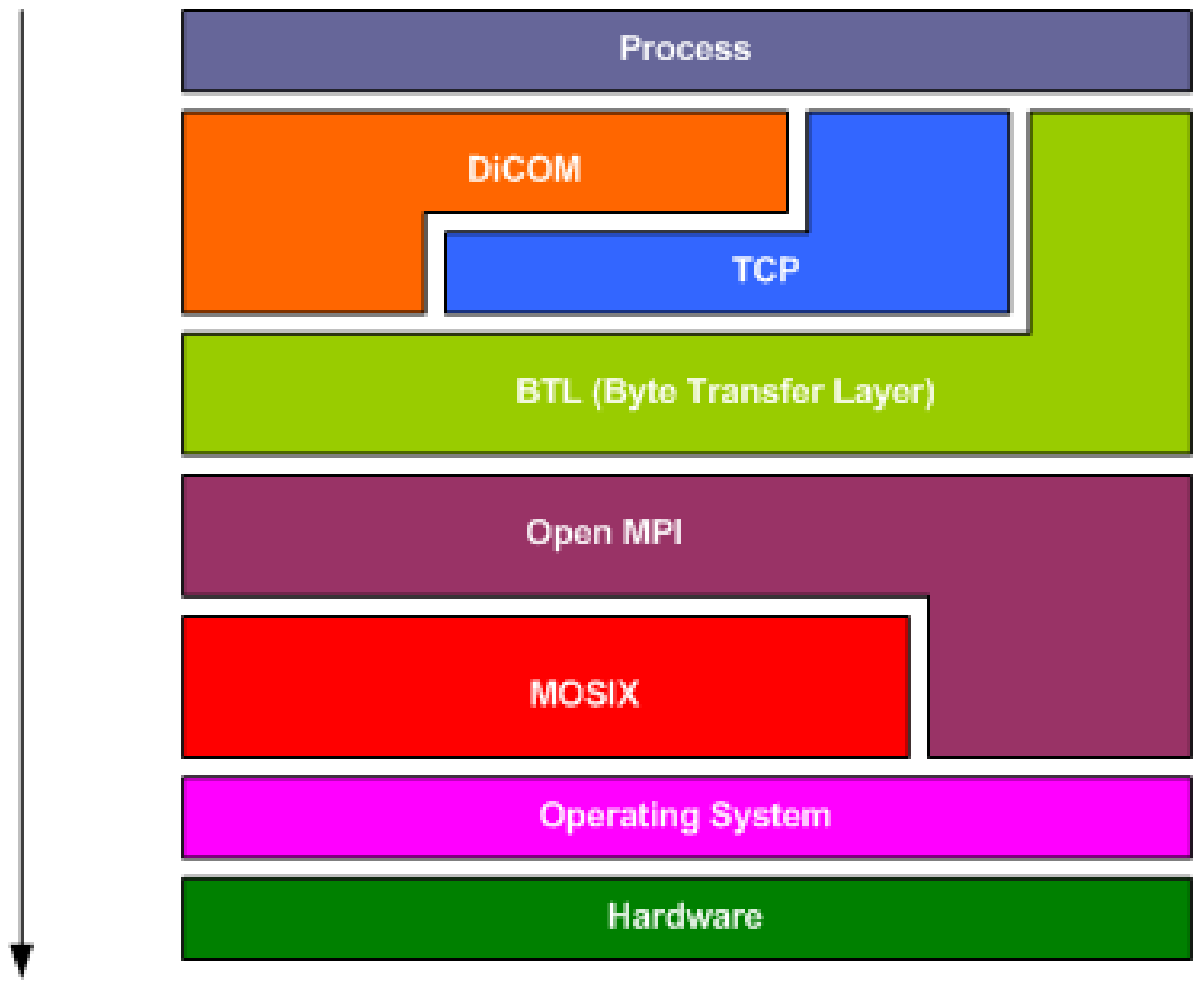

Figure 2.2: The project's layer model
The OMPI relevant modular structure with new MOSIX (DiCOM) addition.

### 2.1.3 Information Dissemination (Gossip) Algorithm

The MOSIX system uses a style of gossip protocol called **Information Dissemination protocol** (rumor-mongering protocols). Which use gossip[2] to spread information (see Figure 2.3); it basically works by flooding agents in the network, but in a manner that produces bounded worst-case loads. The algorithm is used by each and every node to gather information about the state of other nodes in the cluster (load, running processes etc.), this then enables the system to select the optimal manner in which the processes should run (and migrate the appropriate processes).

DiCOM uses the collected data of process locations (running node) to connect directly to the process, bypassing the need to connect via its home node (as depicted in Figure 2.1). Moreover, if DiCOM is unable to determine the "real" location of a process, DiCOM sends the message to the other process's home node directly and, based on the fact that every home node "knows" the location of every job started on it, gets the needed data as a replay and updates its database (see Figure 2.4). Acting in this manner is, effectively, not unlike many common routing algorithms.

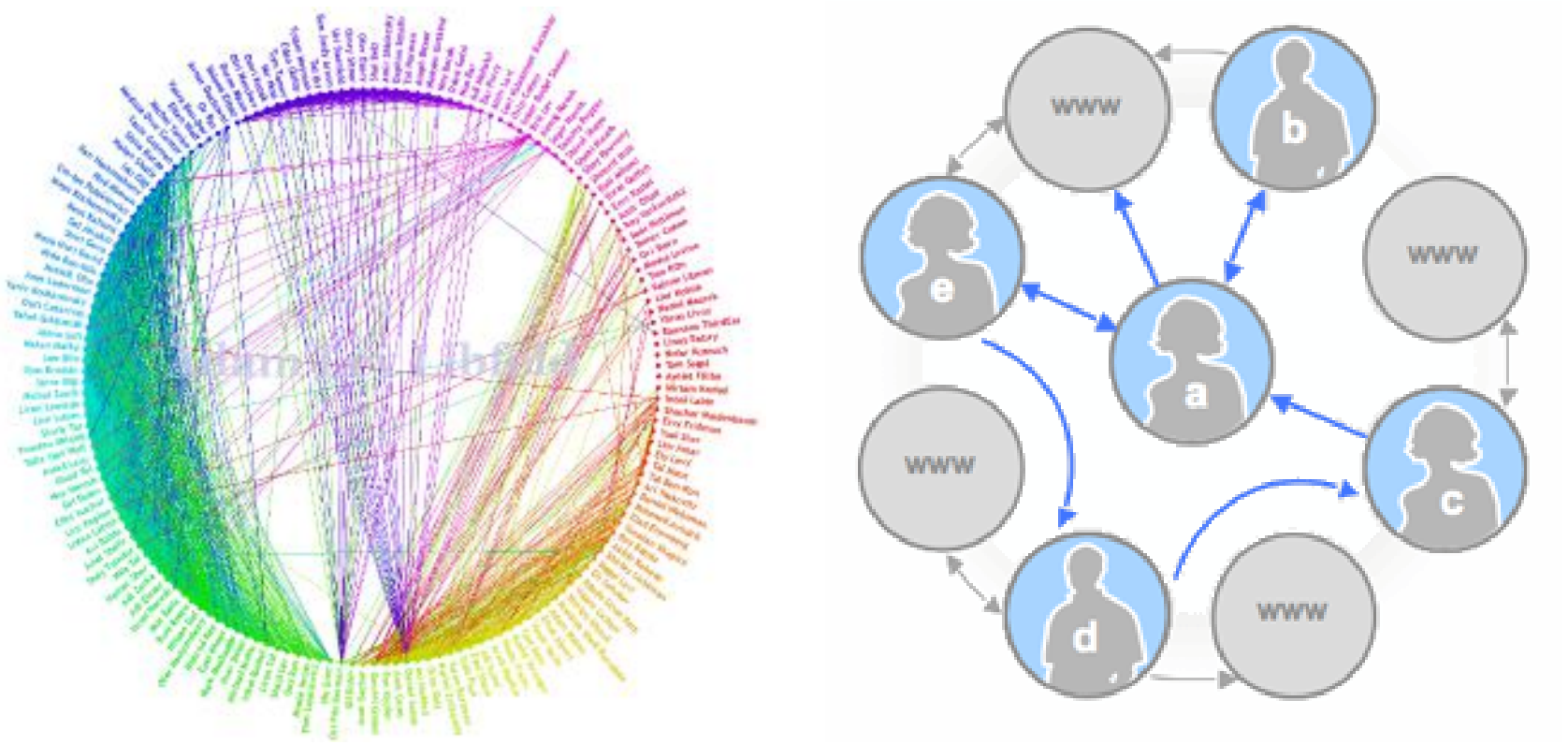

Figure 2.3: Gossip Algorithms
easily tell the world your deepest secrets with some help from your well connected friends.

[2] "Gossip" has the same meaning in both computer and human communication – a bounded (or small), unreliable, semi random transfer of data between pairs of agents who share a common trait (Cluster in computer science, clique in humans etc.). For a more formal description (that more or less reflects that fact) refer to the appropriate appendix section.

### 2.1.4 Other Concepts and Methods

**Mailboxing -** To facilitate DiCOM, each MOSIX process owns a mailbox. This mailbox can contain up to 10,000 unread messages at any given time, up to a total of 32MB. MOSIX Processes can send messages to mailboxes of other processes anywhere within the Cluster or the Multi-Cluster.

**Multi-Channeling -** Much like sockets, a process may want to have more then one channel to his peer. This means that DiCOM must have the ability to locate and filter packets based on the channel through which they came. The implementation includes a unique ID (socket descriptor number) for each such channel, thus, not only does the operating system dispense these unique IDs and "recycle" old ones, but a receiver can immediately match an incoming message to its destination channel.

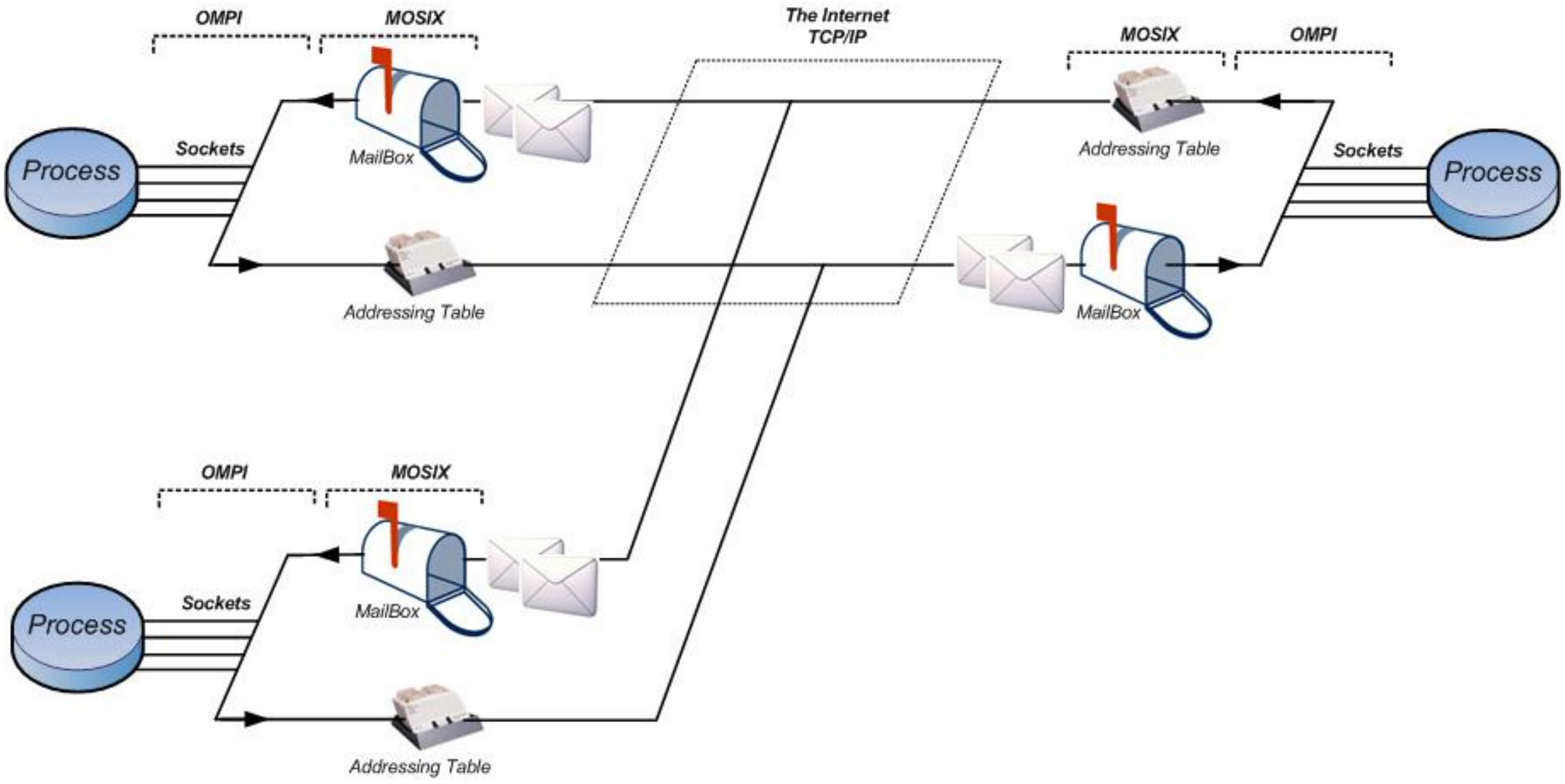

Figure 2.4: Graphic view of the DiCOM protocol
The Mailboxing and addresing concepts

**Transparency -** DiCOM makes the location of processes transparent; so that the sender does not need to know where the receivers run, only to identify them by their home-node and process-ID (PID) in their home-node.

**Consistency -** DiCOM guarantees that the order of messages per receiver is preserved, even when the sender(s) and receiver migrate - no matter where to and how many times they migrate.

**Testing and measurements infrastructure -** For the purpose of testing and evaluating the contribution to the Open-MPI code, extensive testing is to take place and examine the effect in various scenarios. We start with common usage scenarios, move on to examine stability and scalability issues, and finish by looking at extreme conditions such as behavior under heavy traffic loads. A script-based system was created to perform automatic testing, based on an application bank including commonly used MPI software to present relevant results. The system repeats these MPI benchmarks and gathers the result data for future analysis.

## 2.2 Materials, Hardware and Software

### 2.2.1 Used Software

The entire code was written in standard ANSI C, was developed and reviewed using Emacs, and tested on Linux with some Perl scripts. In this project we used the following software resources:

1. Open-MPI source code - Downloaded from openMPI.org, provided with some common usage code examples, no special relation to MOSIX existed. Today the DiCOM module is ready to be submitted.
2. Latest version of MOSIX - Updated several times during the project, including change from 32 bit to 64, some source code was also available. Mosix provides a well build information dissemination daemon ("infod"), and as for today usage of Direct Communication is enabled, but without TCP/IP compatibility (even though it exists and works flawlessly, as can be seen in the results section), DiCOM will be fully integrated into MOSIX as soon as we will finish optimizing it (see future work), with the help of chief MOSIX programmer Amnon Shiloh.
3. CLIP - Cluster management tool.
4. Standard Linux libraries and kernel - Due to the nature of MOSIX, parts of it are compiled in the Linux kernel; therefore, with every change done to the part of DiCOM which is embedded in the kernel, we had to recompile the entire OS.
5. Matlab - Output data analysis and presentation.
6. Virtual Box - virtual machine client.

### 2.2.2 Development and Test Environment

In order to comply with very high compatibility, we developed and tested DiCOM on a verity of environments including: Clusters of raging sizes (2 to 50 nodes), Heterogeneous and homogeneous Clusters, Busy and free Clusters, 32 bit and 64 bit architectures, Intel, SPARC and virtual machines.

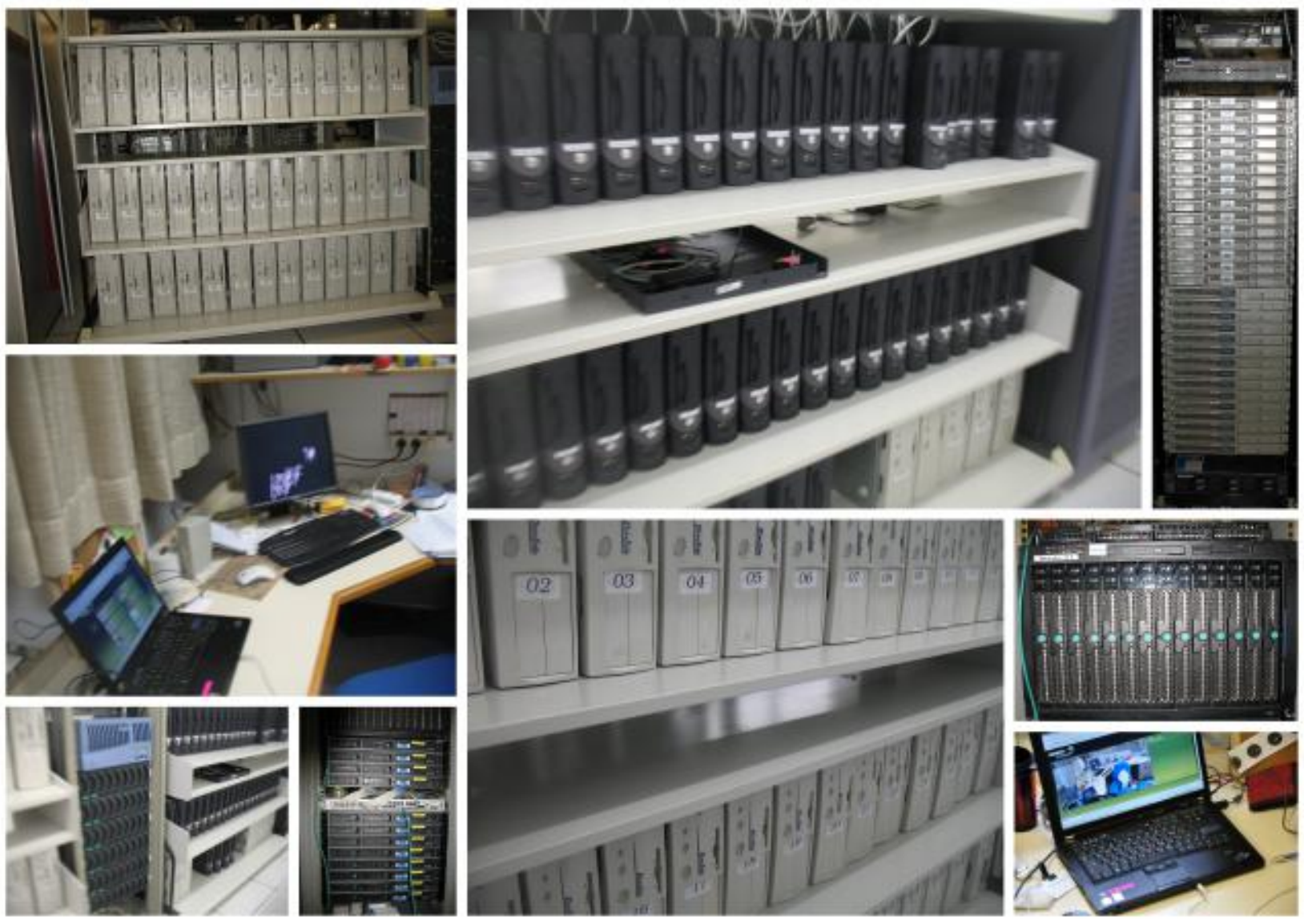

Figure 2.5: Hardware
Some of the hardware used in this project.

# Chapter 3

# Tests and Results

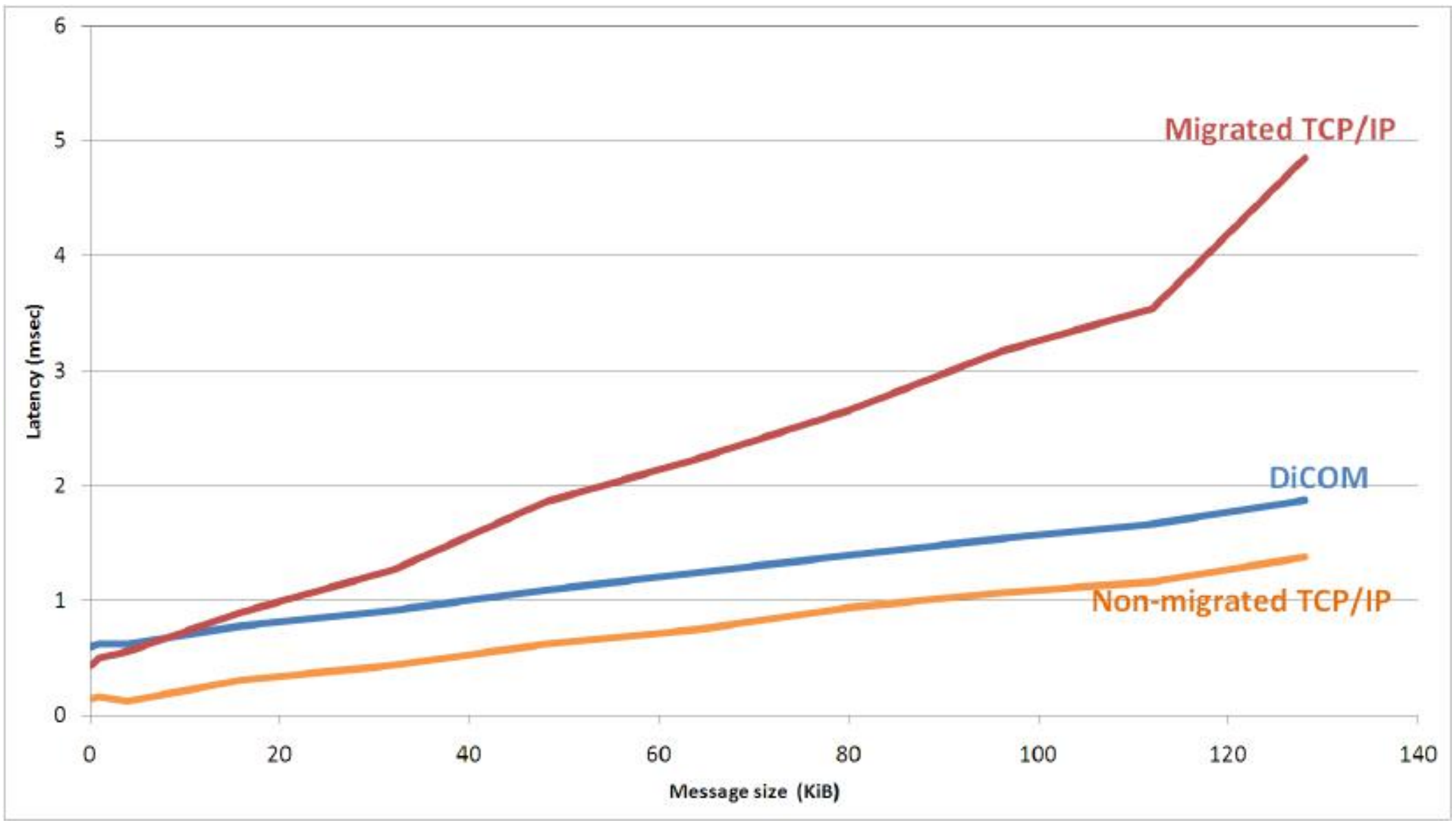

Figure 3.1: Latency test: *small* messages
TCP/IP with and without migration (red and orange respectively) and DiCOM (blue) send/recv latency [mSec] as a function of small message sizes.

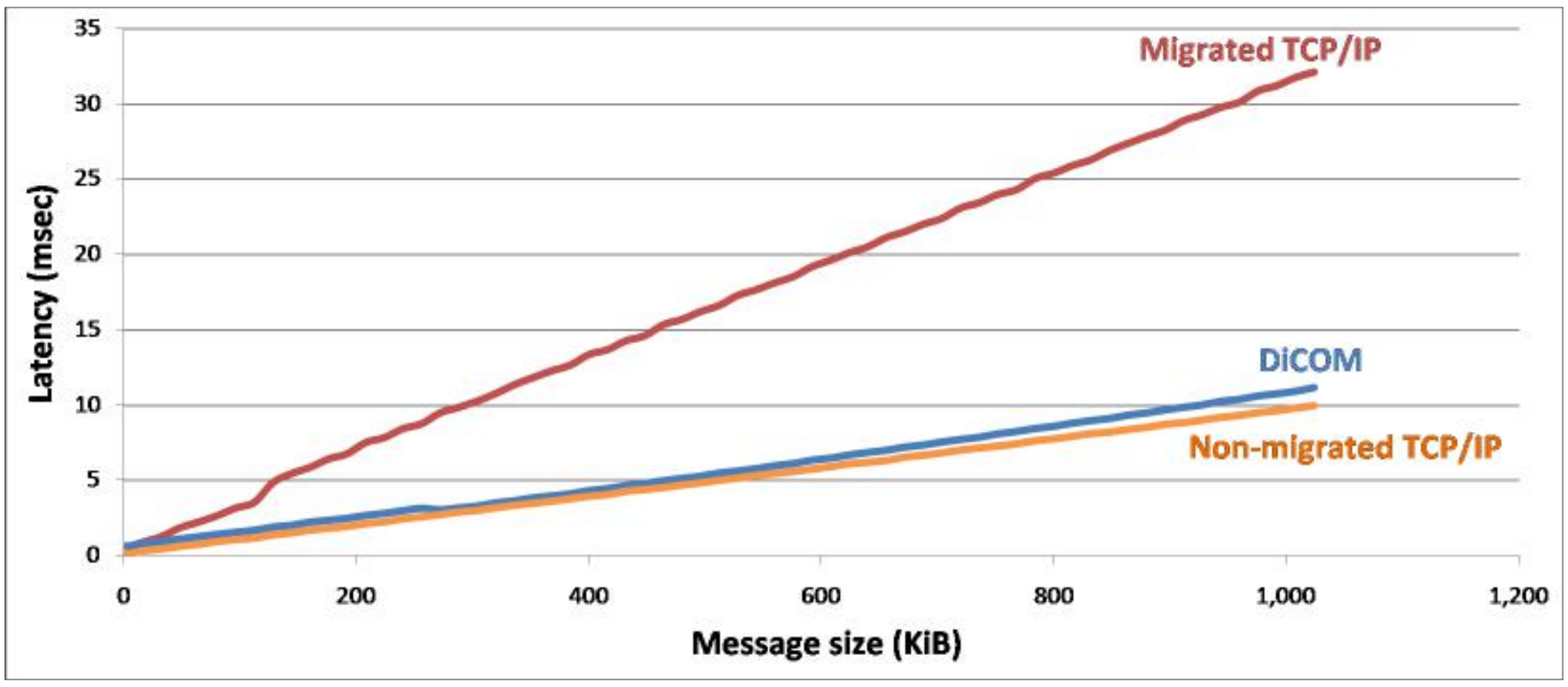

Figure 3.2: Latency test: *typical* messages
TCP/IP with and without migration (red and orange respectively) and DiCOM (blue) send/recv latency [mSec] as a function of typical message size [KiB].

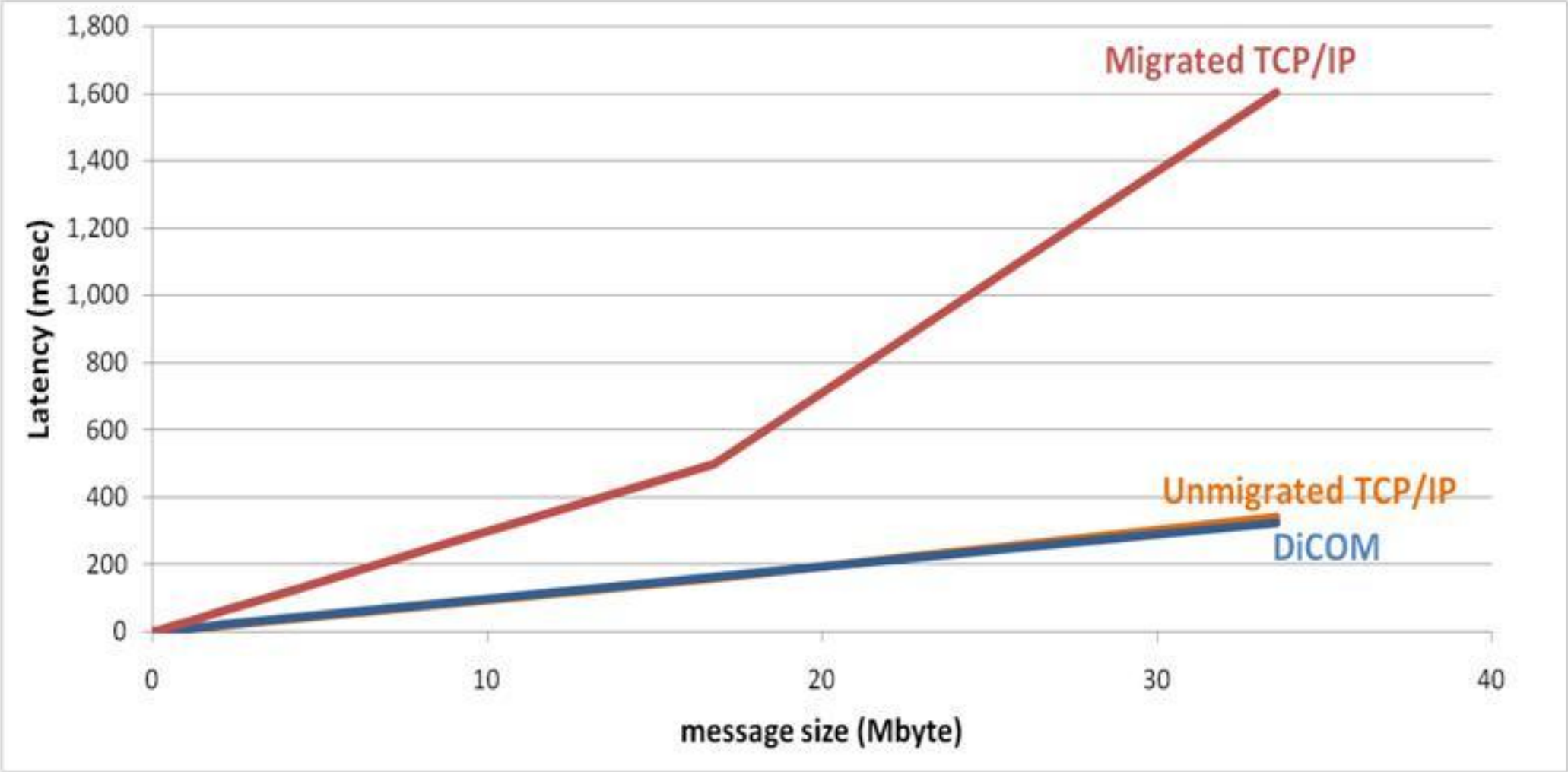

Figure 3.3: Latency test: *large* messages
TCP/IP with and without migration (red and orange respectively) and DiCOM (blue) send/recv latency [mSec] as a function of large message sizes [Mbyte].

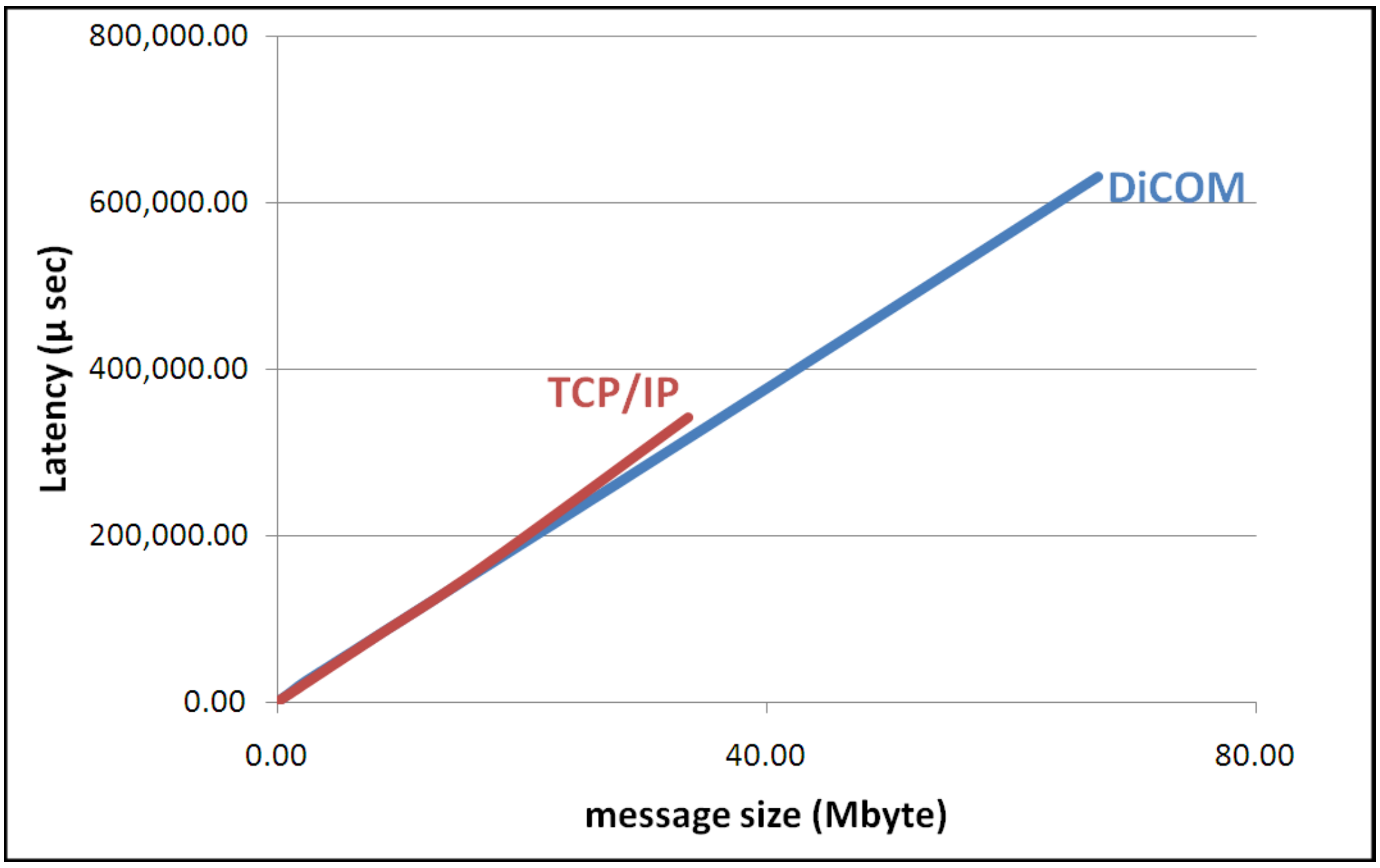

Figure 3.4: Limit test
TCP/IP send/recv latency without migration (red) and DiCOM send/recv latency (blue) [Sec] as a function of large message size [Mbyte].

# Chapter 4

# Conclusions

## 4.1 Result Derived Conclusions

Figures 3.1 to 3.3 present the latencies for different message sizes of TCP/IP with and without process migration vs. DiCOM, which has the same performance regardless of the process locations,
From the figures it can be seen that:

1. The performance of TCP/IP between **non-migrated** process is slightly better then DiCOM for all message sizes. The average slowdown is about **17%** for all message sizes. This is due to the fixed overhead of the DiCOM protocol.
2. DiCOM is slightly slower than **migrated** TCP/IP for small message sizes (below 10k). For larger message sizes, DiCOM is increasingly better than TCP/IP between migrated processes, with an average improvement of **52%** for all of the measured message sizes.

From Figure 3.4, which present the latencies for different message sizes of TCP/IP and DiCOM without process migration, it can be seen that DiCOM can cope with messages twice as large then the maximal message size of TCP/IP.

## 4.2 Design and Implementation Conclusions

1. We have solved problems concerning the integration of two vastly used scientific tools, Open-MPI and MOSIX.
2. We have constructed a tool enabling simple integration of DiCOM into various TCP/IP enabled applications running on MOSIX.
3. **Extensive tests and measurements indicate that the developed tool is stable; it indeed reduces the total runtime of parallel OMPI jobs over MOSIX, and is ready for deployment**.

## 4.3 Future Work

The most significant of possible future developments of this project very well may be the integration of the DiCOM module into the standard MOSIX TCP (see Figure 4.1); this will reduce the work for developer wishing to introduce the system into an existing TCP using code. Moreover, this will enable the addition of a mechanism which will decide dynamically if using TCP will be beneficial in terms of latency, making the TCP/IP latency an upper bound to that of DiCOM.
More on this subject, which may be developed in future work:

1. Parallel computing oriented scheduler built into the MOSIX system.
2. Introduction of the DiCOM module to other parallel computing environments, such as OpenMP and MPICH.
3. Creation socket oriented mailbox, this will enable MOSIX programmers to create a filter free, faster DiCOM module.

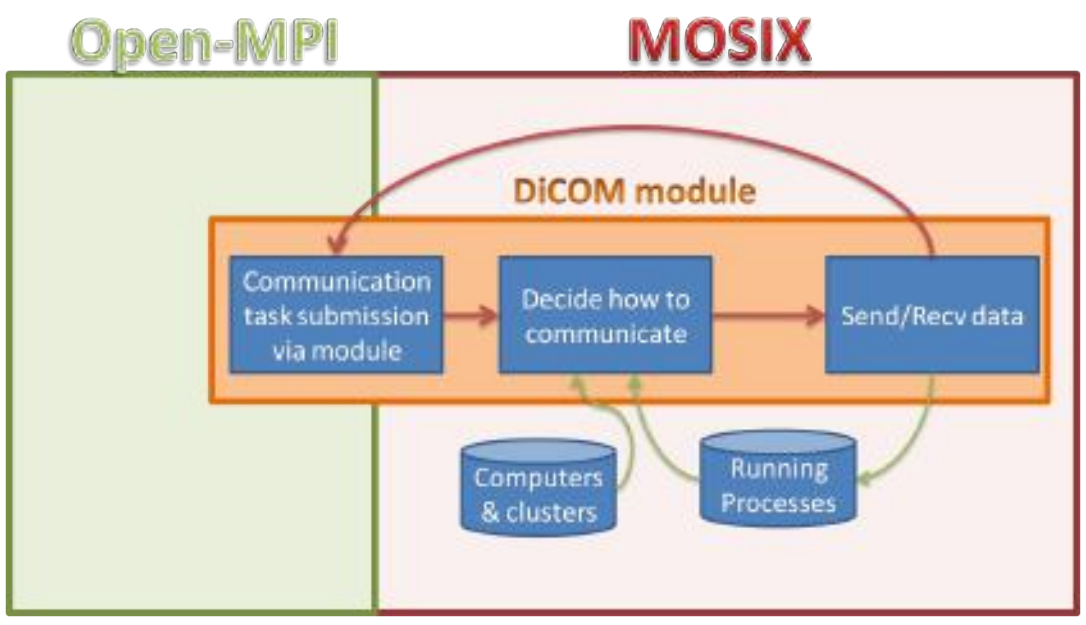

Figure 4.1: Future work - Suggested structure
Suggested structure of mosix embedded DiCOM module with communication manner optimization mechanism.

# Appendix A

# MPI & Open-MPI (OMPI)

MPI (Message Passing Interface) is a language-independent communications protocol used to program parallel computers. Both point-to-point and collective communication are supported. MPI's goals are high performance, scalability, and portability. MPI remains the dominant model used in high-performance computing today. MPI is not sanctioned by any major standards body; nevertheless, it has become a de facto standard for communication among processes that model a parallel program running on a distributed memory system. Actual distributed memory supercomputers such as computer clusters often run these programs.

The Open MPI Project is an open source MPI-2 implementation that is developed and maintained by a consortium of academic, research, and industry partners. Open MPI is therefore able to combine the expertise, technologies, and resources from all across the High Performance Computing community in order to build the best MPI library available.

## A.1 Problems with MPI performance w/o MOSIX

MPI was designed with performance in mind, and as such it contains many optimizations – both at the algorithmic level and the code level. The problem MOSIX can solve is revealed when the workload is imbalanced, causing inefficient allocation of resources (I.e. CPU time) and the undesirable situation when the slowest instance (the result of either a busier or weaker node) is causing the entire job to slow down. The MOSIX system prevents such situations by spreading the computational resources evenly among the MPI instances, thus allowing a fair distribution of resources and better performance for the entire job. Other area influenced by this research is the scheduling of parallel applications. For the time being, the scheduling of a job in the cluster does not take into consideration the current work distribution at the moment. This means that while some of the nodes may already be occupied (busy, yet still accepting new jobs) and others are completely vacant, MPI makes no effort to consider this while assigning the jobs, thus returning us back to the previous issue of load imbalance. With the introduction of the MOSIX system in this area, scheduling looses the dominant effect on runtime due to the load-balancing feature.

## A.2 Problems with running OMPI over MOSIX

The main obstacle with Open-MPI over MOSIX is the use of shared-memory, which is inherently not supported by MOSIX. Since open-MPI merely uses the shared-memory modules as a "shortcut", one can set up open-MPI to avoid the problematic modules – either temporarily by invoking the executables with excluding parameters, or by compiling the open-MPI to exclude these modules entirely. The second obstacle is the establishment of remote nodes, "slaves", as a part of the MPI job. As mentioned previously, Open-MPI typically connects via SSH/RSH and there launches the application instance. If we are to benefit from the use of MOSIX, we must make sure all the instances are run with it. This can be achieved by running the instances locally, where every instance will run over MOSIX like its creator - the original MPI process.
Other issues tackled are the default choice of modules: since MOSIX relays some system-calls over the network, one must optimize the selection of modules to include only the necessary ones, and those which are efficient under these circumstances. For example, interfacing using the local file-system becomes a costly operation over the network, which makes it beneficial to cache data and very inefficient to store intermediate results.

## A.3 Open-MPI modules infrastructure

The Open-MPI source code have a well defined structure and strict naming conventions, the most significant of which is the Modular Component Architecture (MCA), which enables the programmer to add and modify the features of OMPI with great ease:

**MCA:** The Modular Component Architecture (MCA) is the foundation upon which the entire Open MPI project is built. It provides all the component architecture services that the rest of the system uses. Although it is the fundamental heart of the system, it's implementation was designed for HPC – meaning that it is small, fast, and reasonably efficient – and therefore offers few services other finding, loading, and unloading components.

Framework: An MCA framework is a construct that is created for a single, targeted purpose. It provides a public interface that is used by external code, but it also its own internal services. An MCA framework uses the MCA's services to find and load components at run time – implementations of the framework's interface. An easy example framework to discuss is the MPI framework named "BTL", or the Byte Transfer Layer. It is used to sends and receives data on different kinds of networks. Hence, Open MPI has BTL components for shared memory, TCP/IP, Infiniband, MOSIX, Myrinet, etc.

Component: An MCA component is an implementation of a framework's interface. Another common word for component is "plugin." It is a standalone collection of code that can be bundled into a plugin that can be inserted into the Open MPI code base, either at run-time and/or compile-time.

Module: An MCA module is an instance of a component (in the C++ sense of the word "instance"; an MCA component is analogous to a C++ class). For example, if a node running an Open MPI application has multiple Ethernet NICs, the Open MPI application will contain one TCP BTL component, but two TCP BTL modules. This difference between components and modules is important because modules have private state; components do not.

Frameworks, components, and modules can be dynamic or static. That is, they can be available as plugins or they may be compiled statically into libraries (e.g., libmpi).

# Appendix B

# Gossip Protocols and Algorithms

The concept of *gossip communication* can be illustrated by the analogy of office workers spreading rumors. Ted comments to Sally that he believes Fred dyes his mustache. Sally tells Jill, while Ted repeats the idea to Sam. As people move around and repeat the rumor, the number of individuals who have heard the rumor roughly doubles with each "step". The doubling is an approximation that assumes employees communicate at a constant rate, and it doesn't account for gossiping twice to the same person. (Perhaps Ted tries to tell his story to Mark, only to find that Mark already heard it from Jill). Computer systems typically implement this type of protocol with a form of random "peer selection": with a given frequency, each machine picks another machine at random and shares any hot rumors.
The power of gossip lies in the robust spread of information. Even if Jill had trouble understanding Sally (perhaps she was whispering), she will probably run into someone else soon and can learn the news that way.
Expressing these ideas in more technical terms, a gossip protocol is one that satisfies the following conditions:

1. The core of the protocol involves periodic, paired, inter-process interactions.
2. The information exchanged during these interactions is of bounded size.
3. When agents interact, the state of at least one agent changes to reflect the state of the other. A gossip interaction does not occur when A pings B just to measure the response time, as this does not involve the transmittal of state between agents.
4. Reliable communication is not assumed.
5. The frequency of the interactions is low compared to typical message latencies so that the protocol costs are negligible.
6. There is some form of randomness in the peer selection. Peers might be selected from the full set of nodes or from a smaller set of "neighbors".

As mention before, MOSIX is using a specific gossip protocol: **Dissemination protocols** (or rumor-mongering protocols) - These use gossip to spread information; they basically work by flooding agents in the network, but in a manner that produces bounded worst-case loads:
*Event dissemination protocols* use gossip to carry out multicasts. They report events, but the gossip occurs periodically and events don't actually trigger the gossip. One concern here is the potentially high latency from when the event occurs until it is delivered.
*Background data dissemination protocols* continuously gossip about information associated with the participating nodes. Typically, propagation latency isn't a concern, perhaps because the information in question changes slowly or there is no significant penalty for acting upon slightly stale data.

# Appendix C

# Acronym Dictionary **or everything nowadays have a TLA**

ACK - Acknowledgement
BTL - Byte Transfer Layer
COW - Cluster Of Workstations
CPU - Central Processing Unit
DiCOM - Direct Communication Over MOSIX
HPC - High Performance Computing
I/O - Input/Output
LOBOS - Lots Of Boxes On Shelves
MCA - Modular Component Architecture
MPI - Message Passing Interface
NIC - Network Interface Controller
NOW - Network Of Workstations, sometimes called COW
MSG - Message, sometimes (not in this work) Mono Sodium Glutamate
OMPI - OpenMPI - Open Message Passing Interface
OpenMP - Open Multi-Processing
OS - Operating System
PID - Process ID
RTE - RunTime Environment
TCP/IP - Transmission Control Protocol
TLA - Three Letter Acronym
RSH - Remote Shell
SSH - Secure Shell

# References and other relevant literature

# Index